\documentclass[12pt,aps,prd,preprint,tightenlines,
   showpacs,nofootinbib]{revtex4}
\newcommand{\PRE}[1]{{#1}} 

\usepackage{bm} \usepackage{epsfig}

\newcommand{\gweak}{g_{\text{weak}}}
\newcommand{\mweak}{m_{\text{weak}}}
\newcommand{\mplanck}{M_{\text{Pl}}}

\newcommand{\ev}{\text{eV}}
\newcommand{\kev}{\text{keV}}
\newcommand{\mev}{\text{MeV}}
\newcommand{\gev}{\text{GeV}}
\newcommand{\tev}{\text{TeV}}

\newcommand{\s}{\text{s}}

\newcommand{\eg}{{\em e.g.}}

\newcommand{\eqref}[1]{Eq.~(\ref{#1})}
\newcommand{\eqsref}[2]{Eqs.~(\ref{#1}) and (\ref{#2})}
\newcommand{\secref}[1]{Sec.~\ref{sec:#1}}

\newcommand{\figref}[1]{Fig.~\ref{fig:#1}}
\newcommand{\figsref}[2]{Figs.~\ref{fig:#1} and \ref{fig:#2}}

\newcommand{\stau}{\tilde{\tau}}

\newcommand{\be}{\begin{equation}}
\newcommand{\ee}{\end{equation}}

\newcommand{\tBBN}{t_{\text{BBN}}}
\newcommand{\TBBN}{T_{\text{BBN}}}
\newcommand{\ThBBN}{T^h_{\text{BBN}}}
\newcommand{\tRH}{t_{\text{RH}}}
\newcommand{\TRH}{T_{\text{RH}}}
\newcommand{\xiRH}{\xi_{\text{RH}}}
\newcommand{\ThRH}{T^h_{\text{RH}}}
\newcommand{\TCMB}{T_{\text{CMB}}}
\newcommand{\ThCMB}{T^h_{\text{CMB}}}
\newcommand{\glight}{g_{\text{light}}}
\newcommand{\gheavy}{g_{\text{heavy}}}
\newcommand{\ghlightBBN}{g^{h\, \text{BBN}}_{\text{light}}}
\newcommand{\ghheavyBBN}{g^{h\, \text{BBN}}_{\text{heavy}}}
\newcommand{\ghlightCMB}{g^{h\, \text{CMB}}_{\text{light}}}
\newcommand{\ghheavyCMB}{g^{h\, \text{CMB}}_{\text{heavy}}}
\newcommand{\neff}{N_{\text{eff}}}

\begin{document}

\preprint{UCI-TR-2008-26}

\title{
\PRE{\vspace*{1.5in}}
Thermal Relics in Hidden Sectors
\PRE{\vspace*{0.3in}}
}

\author{Jonathan L.~Feng}
\affiliation{Department of Physics and Astronomy, University of
California, Irvine, CA 92697, USA
\PRE{\vspace*{.5in}}
}

\author{Huitzu Tu%
\PRE{\vspace*{.2in}}
}
\affiliation{Department of Physics and Astronomy, University of
California, Irvine, CA 92697, USA
\PRE{\vspace*{.5in}}
}

\author{Hai-Bo Yu%
\PRE{\vspace*{.2in}}
}
\affiliation{Department of Physics and Astronomy, University of
California, Irvine, CA 92697, USA
\PRE{\vspace*{.5in}}
}

\begin{abstract}
\PRE{\vspace*{.3in}}  Dark matter may be hidden, with no standard
model gauge interactions. At the same time, in WIMPless models with
hidden matter masses proportional to hidden gauge couplings squared,
the hidden dark matter's thermal relic density may naturally be in
the right range, preserving the key quantitative virtue of WIMPs. We
consider this possibility in detail. We first determine
model-independent constraints on hidden sectors from Big Bang
nucleosynthesis and the cosmic microwave background. Contrary to
conventional wisdom, large hidden sectors are easily accommodated. A
flavour-free version of the standard model is allowed if the hidden
sector is just 30\% colder than the observable sector after
reheating. Alternatively, if the hidden sector contains a
1-generation version of the standard model with characteristic mass
scale below 1 MeV, even identical reheating temperatures are
allowed. We then analyze hidden sector freezeout in detail for a
concrete model, solving the Boltzmann equation numerically and
understanding the results from both observable and hidden sector
points of view. We find that WIMPless dark matter indeed obtains the
correct relic density for masses in the range $\kev \lesssim m_X
\lesssim \tev$. The upper bound results from the requirement of
perturbativity, and the lower bound assumes that the observable and
hidden sectors reheat to the same temperature and is raised to the
MeV scale if the hidden sector is 10 times colder. WIMPless dark
matter therefore generalizes the WIMP paradigm to the largest mass
range possible for viable thermal relics and provides a unified
framework for exploring dark matter signals across nine orders of
magnitude in dark matter mass.
\end{abstract}


\maketitle

\section{Introduction}
\label{sec:intro}

At present, all solid evidence for dark matter is gravitational.  At
the same time, the possibility that dark matter has electromagnetic or
strong interactions is highly
constrained~\cite{Dimopoulos:1989hk,Starkman:1990nj}.  A
straightforward possibility, then, is that dark matter is hidden,
consisting of particles that have no standard model (SM) gauge
interactions.

Hidden sectors have a long and distinguished history.  For example,
the idea that a hidden sector may restore parity on cosmological
scales is as old as the idea of parity violation
itself~\cite{Lee:1956qn}.  Such hidden sectors, containing ``mirror
matter,'' have been studied by many groups, as have their possible
implications for dark matter~\cite{Blinnikov:1982eh,Hodges:1993yb,%
Berezhiani:1995am,Mohapatra:2000qx,Berezhiani:2000gw,Mohapatra:2001sx,%
Ignatiev:2003js,Foot:2003iv,Berezhiani:2003wj,Foot:2004wz}.  String
theory also motivates hidden sectors with their own gauge
interactions~\cite{Green:1984sg}, as in the case of the heterotic
string~\cite{Gross:1984dd} and intersecting brane
models~\cite{Blumenhagen:2005mu}.  More recently, hidden sectors have
been central to several phenomenological developments, including Higgs
portals~\cite{Schabinger:2005ei,Patt:2006fw}, hidden
valleys~\cite{Strassler:2006im}, unparticles~\cite{Georgi:2007ek}, and
quirks~\cite{Kang:2008ea}, and the possibility that dark matter may
reside in a hidden sector has been discussed in several recent
works~\cite{Chen:2006ni,Kikuchi:2007az,MarchRussell:2008yu,%
Hooper:2008im,McDonald:2008up,Kim:2008pp,Krolikowski:2008qa,%
Gong:2008uz,Feng:2008ya,Foot:2008nw}.

In considering hidden dark matter, one would seemingly be sacrificing
critical virtues of more conventional dark matter candidates, namely,
the connection of dark matter to the gauge hierarchy problem, and the
fact that weakly-interacting massive particles (WIMPs) naturally have
the desired thermal relic density.  This is not necessarily true,
however.  In the recently proposed framework of WIMPless dark
matter~\cite{Feng:2008ya}, hidden sector dark matter also naturally
has the desired thermal relic density.  The essential idea is that,
very generally, a dark matter candidate that decoupled from the
thermal bath non-relativistically has a thermal relic density
\begin{equation}
\Omega h^2 \sim \frac{1}{\langle \sigma_A v \rangle }
\sim \frac{m^2}{g^4} \ ,
\label{omega}
\end{equation}
where $\langle \sigma_A v \rangle$ is the thermally-averaged
annihilation cross section times velocity, and $g$ and $m$ are the
dark matter particle's gauge coupling and mass.  For WIMPs with weak
interaction coupling constant $\gweak \simeq 0.65$ and weak-scale mass
$\mweak \sim 100~\gev - 1~\tev$, this relic mass density is naturally
near the observed $\Omega_{\text{DM}} h^2 \simeq 0.11$.

In WIMPless models, hidden sector particles $X$ have masses and
couplings that may be very different from WIMPs, but which
nevertheless satisfy
\begin{equation}
\frac{m_X}{g_X^2} \sim \frac{\mweak}{\gweak^2} \ .
\label{scaling}
\end{equation}
In the WIMPless examples discussed in Ref.~\cite{Feng:2008ya},
\eqref{scaling} follows from the structure of gauge-mediated
supersymmetry breaking (GMSB), which generates hidden sector masses
that are proportional to couplings squared.  The WIMPless framework
therefore naturally generalizes the WIMP paradigm to other cold dark
matter candidates without sacrificing the key thermal relic density
virtue of WIMPs.  If additional connector particles with both SM and
hidden gauge quantum numbers are present, they will mediate SM-hidden
interactions.  The WIMPless framework may therefore also predict
qualitatively new signals, such as an explanation of the DAMA signal
in terms of GeV dark matter~\cite{Feng:2008dz}.

In this paper, we study hidden thermal relics in general, and WIMPless
dark matter in particular.  For the most part, we assume that there
are no connector particles, so that the hidden sector interacts with
the SM only very weakly.  In \secref{BBN}, we carry out a
model-independent analysis of constraints on hidden sectors from Big
Bang nucleosynthesis (BBN), the cosmic microwave background (CMB), and
other cosmological data.  As is well known, these exclude the
possibility that the hidden sector is an exact copy of the
SM~\cite{Kolb:1985bf}.  This statement assumes that the hidden and
observable sectors are at the same temperature at late times, however.
In the present context, there is no strong reason to assume that the
observable and hidden sectors are reheated to the same
temperature~\cite{Hodges:1993yb,Berezhiani:1995am}, and, even if they
are, for a truly decoupled hidden sector, their temperatures will
differ by the time of BBN.  Following previous works in the context of
mirror matter (see, \eg, Refs.~\cite{Hodges:1993yb,Berezhiani:1995am,%
Foot:1996hp,Berezhiani:2000gw,Ignatiev:2003js}), we consider the
cosmological constraints allowing different temperatures and find that
even large hidden sectors are easily accommodated. For example, a
hidden version of the SM is allowed even if the hidden sector's
reheating temperature is as much as 70\% of the observable sector's.
Alternatively, if the hidden sector contains a 1-generation version of
the MSSM with characteristic mass scale below 1 MeV, even identical
reheating temperatures are allowed.

We then present a concrete model for the hidden sector in
\secref{concrete} and study the freezeout of dark matter in this
hidden sector in \secref{relic}.  The relations of
\eqsref{omega}{scaling} are, of course, only rough relations capturing
the parametric dependences.  In these Sections, we present a detailed
analysis of a specific case that highlights several subtleties and
generalizes previous freezeout analyses to cases with sectors at
different temperatures.  This analysis confirms the validity of
conclusions based on \eqsref{omega}{scaling} and highlights various
subtleties of the WIMPless framework.  In particular, we find that
WIMPless dark matter may have the desired thermal relic density for
the entire range of $\kev \alt m_X \alt \tev$, where the lower bound
is set by the requirement that dark matter freezeout is
non-relativistic, and the upper bound follows from the requirement of
perturbative gauge couplings. WIMPless dark matter therefore
encompasses as large a range of masses as one could expect of dark
matter that has the naturally correct thermal relic density, and it
provides a unified framework for addressing many diverse dark matter
signals and phenomenology.

Finally, in \secref{adding}, we discuss modifications from the
presence of connector particles with both hidden and observable sector
gauge interactions.  We find that, under general assumptions, these
fields do not upset the conclusions derived earlier.  We present our
conclusions in \secref{conclusions}.

\section{BBN and CMB Constraints}
\label{sec:BBN}

We assume that the observable sector is the minimal supersymmetric
standard model (MSSM), which is supplemented by a single hidden
sector.  Hidden sector parameters are denoted by the superscript $h$.
For most of this work, we assume that there are no connectors, fields
with both SM and hidden gauge charges. The observable and hidden
sectors therefore interact extremely weakly with each other, and the
observable sector's temperature $T$ need not equal the hidden sector's
temperature $T^h$.  We define
\begin{equation}
\xi(t) = \frac{T^h(t)}{T(t)}
\end{equation}
to parametrize the mismatch.  In addition, it will be convenient to
introduce the standard relativistic degrees of freedom parameters.
Letting
\begin{equation}
C_i = \left\{ \begin{array}{l}
1 , \ i = \text{boson} \\
\frac{7}{8} , \ i = \text{fermion} \ , \end{array} \right.
\end{equation}
we define
\begin{equation}
g_* (T) \equiv
\sum_{m_i < T} C_i \, g_i \left( \frac{T_i}{T} \right)^4 \qquad
g_{*S} (T) \equiv
\sum_{m_i < T} C_i \, g_i \left( \frac{T_i}{T} \right)^3 \ ,
\label{obsg}
\end{equation}
where the sum is over observable sector particles, and
\begin{equation}
g_*^h (T^h) \equiv
\sum_{m_i < T^h} C_i \, g_i \left( \frac{T_i}{T^h} \right)^4 \qquad
g_{*S}^h (T^h) \equiv
\sum_{m_i < T^h} C_i \, g_i \left( \frac{T_i}{T^h} \right)^3
\label{hiddeng}
\end{equation}
where the sum is over hidden sector particles.  $T_i$ is the
temperature of particle $i$, and $g_i$ denotes its internal degrees of
freedom.  Throughout this work, we assume that the gravitino has a
negligible relic density, as is the case, for example, for low and
moderate reheating temperatures, and the gravitino is not included
\eqsref{obsg}{hiddeng}.

BBN is sensitive to the expansion rate of the Universe at time $\tBBN
\sim 1~\s$ and temperature $\TBBN \sim 1~\mev$.  The expansion rate is
determined by the energy density, and so constrains light particles
even if they have no SM
interactions~\cite{Steigman:1977kc,Steigman:2007xt}. The constraint is
conventionally quoted as a bound on $\neff$, the effective number of
light neutrino species, and may be taken to be $\neff = 3.24 \pm 1.2$
(95\% CL), where the baryon density has been fixed to the value
determined by the CMB, and both $^4$He and D data are
included~\cite{Cyburt:2004yc,Fields:2006ga}.  In our framework, this
implies
\begin{equation}
g_*^h(\ThBBN)
\left(\frac{\ThBBN}{\TBBN} \right)^4 =\frac{7}{8}\, \cdot 2 \cdot
(\neff - 3) \le 2.52 \ (\text{95\% CL}) \ ,
\label{bound}
\end{equation}
where $\ThBBN$ is the temperature of the hidden sector at time
$\tBBN$.  If the hidden sector is an exact copy of the MSSM and
$\ThBBN = \TBBN$, $g_*^h(\ThBBN) = 10.75$.  This violates the bound of
\eqref{bound}, as is well-known~\cite{Kolb:1985bf}, and may create the
impression that large hidden sectors of the size of the MSSM are
completely excluded.

If the observable and hidden sectors are not in thermal contact,
however, the hidden sector may be colder than the observable sector.
This would be the case if, for example, the inflaton couplings to the
observable and hidden sectors are not identical, so that they reheat
to different temperatures~\cite{Hodges:1993yb,Berezhiani:1995am}.
Alternatively, the observable and hidden sectors may initially have
the same temperature, either because they have the same inflaton
couplings or because they are in thermal contact, but may cool
independently and have different temperatures at later times.  In
fact, it quite reasonable that the observable and hidden sectors lose
thermal contact before BBN.  Thermal contact would require efficient
interactions between the two sectors at low temperatures.  These would
require light connector fields to mediate these interactions, but
light fields with SM gauge charges generically violate experimental
constraints.  We conclude, then, that it is far from guaranteed that
the observable and hidden sectors will be at the same temperature at
$t_{\text{BBN}}$; generically, they will be at different temperatures.

To explore this possibility, we assume that the observable and hidden
sectors reheat at time $\tRH$ to temperatures $\TRH$ and $\ThRH$,
respectively.  We further assume that the entropy per comoving volume
is conserved in each sector separately, so that
\begin{equation}
\frac{g_{*S}^h(\ThBBN) T^{h\, 3}_{\text{BBN}}}
     {g_{*S}^h(\ThRH) T^{h\, 3}_{\text{RH}}} =
\frac{g_{*S}(\TBBN) \TBBN^3}{g_{*S}(\TRH) \TRH^3} \ .
\label{entropy}
\end{equation}
Given mass spectra in the observable and hidden sectors, reheat
temperatures $\TRH$ and $\ThRH$, and \eqref{entropy}, the ratio
$\ThBBN/\TBBN$ is fixed, and one can determine if \eqref{bound} is
satisfied.

To provide numerical examples, we define
\begin{equation}
\xiRH \equiv \frac{\ThRH}{\TRH} \ ,
\end{equation}
the ratio of temperatures just after reheating, and assume $\TRH$ is
above the mass of all MSSM particles.  We also define
\begin{equation}
\glight \equiv \sum_{m_i < \TBBN} \! \! \! \! \! C_i \, g_i \qquad
\gheavy \equiv \sum_{\TBBN < m_i < \TRH} \! \! \! \! \! C_i \, g_i
\end{equation}
for the observable sector and
\begin{equation}
\label{gh_heavy_light}
\ghlightBBN \equiv \sum_{m_i < \ThBBN} \! \! \! \! \! C_i \, g_i \qquad
\ghheavyBBN \equiv \sum_{\ThBBN < m_i < \ThRH} \! \! \! \! \! C_i \, g_i
\end{equation}
for the hidden sector, where we have assumed all hidden relativistic
degrees of freedom have a common temperature.  With these definitions,
\eqref{entropy} becomes
\begin{equation}
\frac{\ghlightBBN}{\ghlightBBN + \ghheavyBBN}
\left(\frac{\ThBBN}{\TBBN}\right)^3 =
\frac{\glight}{\glight + \gheavy} \xiRH^3 \ .
\end{equation}
Solving for $\ThBBN/\TBBN$ and substituting the MSSM values
$\glight = 10.75$ and $\glight + \gheavy = 228.75$, we find that
the BBN bound of \eqref{bound} becomes
\begin{equation}
\ghheavyBBN \le \left( \frac{3.49}{\xiRH} \right)^3
\left( \ghlightBBN \right)^{\frac{1}{4}} - \ghlightBBN \ .
\label{bound2}
\end{equation}

\begin{figure}[tb]
\begin{center}
\includegraphics*[width=12cm,clip=]{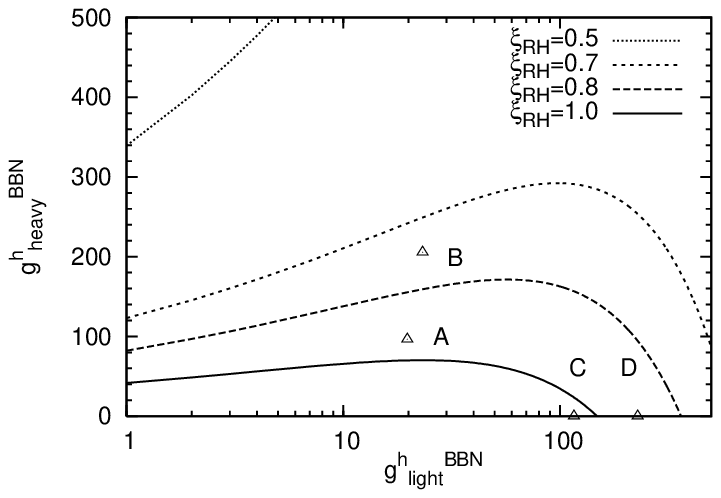}
\end{center}
\vspace*{-.29in}
\caption{Bounds from BBN in the $(\ghlightBBN, \ghheavyBBN)$ plane,
where $\ghlightBBN$ and $\ghheavyBBN$ are the hidden degrees of
freedom with masses $m < \ThBBN$ and $\ThBBN < m < \ThRH$,
respectively, for $\xiRH \equiv \ThRH/\TRH = 0.5$, 0.7, 0.8, 1.0
(from top to bottom).  The regions above the contours are excluded.
We assume that the observable sector reheats to a temperature above
the mass of all MSSM particles.  The values of $(\ghlightBBN,
\ghheavyBBN)$ are marked for four example hidden sectors: (A)
1-generation and (B) 3-generation flavor-free versions of the MSSM
with $\ThBBN < m_X < \ThRH$, and (C) 1-generation and (D)
3-generation flavor-free versions of the MSSM with $m_X < \ThBBN/2$
(see \secref{concrete}). } \label{fig:gstar}
\end{figure}

Bounds in the $(\ghlightBBN, \ghheavyBBN)$ plane for various values of
$\xiRH$ are given in \figref{gstar}. Several aspects deserve comment.
As evident from \eqref{bound2}, the bound is extremely sensitive to
$\xiRH$. The $\xiRH = 0.5$ contour is barely contained on the plot,
implying that if the hidden sector reheats to half the temperature of
the observable sector, very large hidden sectors are allowed.  Several
example hidden sectors, defined precisely in \secref{concrete}, are
also given on the plot.  Model A is a 1-generation flavor-free version
of the MSSM, with all Yukawa couplings of order 1.  We see that it is
allowed for $\xiRH < 0.92$, that is, reheat temperatures
that are almost identical to the observable sector. This is our
prototype model, which will be used in our detailed freezeout analysis
in \secref{relic}. Model B is a full 3-generation flavor-free version
of the MSSM. We see that even a hidden sector as ``large'' as the MSSM
is allowed, provided $\xiRH < 0.74$.

Perhaps even more interesting, \figref{gstar} shows that Model C, a
1-generation version of the MSSM with all degrees of freedom
relativistic at BBN, is allowed, even for $\xiRH =1.06$.  Naively, one
might expect a hidden sector that starts at the same temperature as
the MSSM and has over 100 light degrees of freedom to be completely
excluded.  In fact, however, such a hidden sector is much colder than
the MSSM at BBN times, since its cooling is not slowed by the
disappearance of heavy degrees of freedom, in contrast to the MSSM.
This cooling is critical, given the 4th power of the temperature in
\eqref{bound}, and makes such a hidden sector allowed.  For similar
reasons, \figref{gstar} shows that an excluded hidden sector is only
made more excluded by the addition of heavy degrees of freedom.  Note,
however, that \figref{gstar} also shows that in some cases, rather
counter-intuitively, an excluded hidden sector may be made allowed by
the addition of {\em light} degrees of freedom.

CMB temperature anisotropy measurements provide another constraint on
the number of relativistic species.  WMAP 5 year data, combined with
distance information from baryon acoustic oscillations, supernovae
(SNIa) and Hubble constant measurements, imply $\neff = 4.4 \pm 1.5$
(68\% CL)~\cite{Komatsu:2008hk}.  In contrast to the BBN bound, this
is a 68\% CL bound; at 95\% CL, the bound is very weak.  This bound is
therefore numerically much weaker than the BBN bound.  In the future,
however, CMB data from, for example, the Planck
satellite~\cite{:2006uk} will achieve a much higher sensitivity in
constraining the amount of radiation
energy~\cite{Hannestad:2006as,Hamann:2007pi,Ichikawa:2008pz,Popa:2008tb}.
At the same time, of course, the CMB bound constrains the energy
density at much later times with redshift $z \sim 1000 - 3000$ and
observable sector temperatures $T \sim 1~\ev$.  It is thus interesting
to explore the CMB constraint when the hidden sector includes
particles with mass between 1 MeV and 1 eV.  (See also
Ref.~\cite{Simha:2008zj} for a combined constraint.)

The current CMB constraint in our framework is
\begin{equation}
g^h_\ast (\ThCMB)\, \left(\frac{\ThCMB}{\TCMB}\right)^4
= \frac{7}{8}\, \cdot 2\, \cdot\, (\neff - 3.046)\,
\left(\frac{T_\nu}{T_\gamma}\right)^4 \leq 1.30 \ \text{(68\% CL)} \, ,
\end{equation}
where $T_\nu / T_\gamma = (4/11)^{1/3}$ is the neutrino to photon
temperature ratio after the era of $e^\pm$ annihilation. The fact that
the SM neutrinos were not fully decoupled at $e^\pm$ annihilation is
taken into account by using the modified number of SM neutrinos of
3.046 instead of 3~\cite{Mangano:2005cc,Coleman:2003hs}.  Following
the BBN analysis, this CMB bound implies
\begin{equation}
\ghheavyCMB \leq \left(\frac{4.14}{\xiRH} \right)^3
  \left( \ghlightCMB \right) ^{\frac{1}{4}} - \ghlightCMB \ ,
\end{equation}
where $\ghheavyCMB$ and $\ghlightCMB$ are defined as in
\eqref{gh_heavy_light}, but with $\ThBBN$ replaced by $\ThCMB$.

\begin{figure}[tb]
\begin{center}
\includegraphics*[width=12cm,clip=]{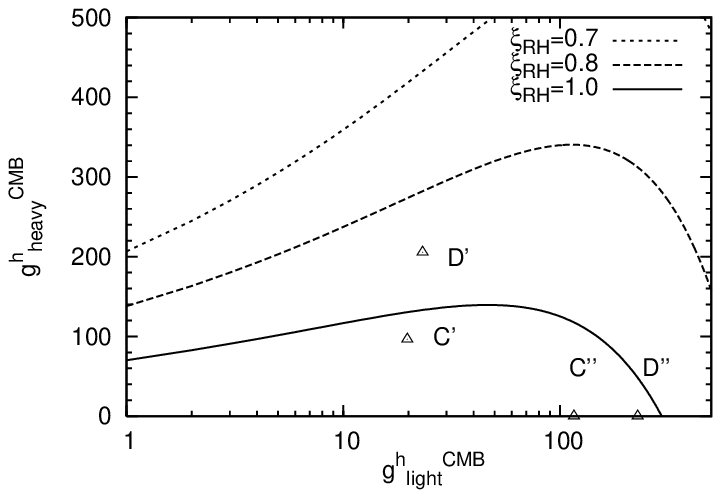}
\end{center}
\vspace*{-.29in}
\caption{As in \figref{gstar}, but for bounds from the CMB in the
$(\ghlightCMB, \ghheavyCMB)$ plane, where $\ghlightCMB$ and
$\ghheavyCMB$ are the hidden effective degrees of freedom with
masses $m < \ThCMB$ and $\ThCMB < m < \ThRH$, respectively.  The
values of $(\ghlightCMB, \ghheavyCMB)$ are given for four example
hidden sectors: (C$^\prime$) 1-generation and (D$^\prime$)
3-generation flavor-free versions of the MSSM with $\ThCMB < m_X <
\ThRH$, and (C$^{\prime\prime}$) 1-generation and
(D$^{\prime\prime}$) 3-generation flavor-free versions of the MSSM
with $m_X < \ThCMB/2$ (see \secref{concrete}).  }
\label{fig:gstarcmb}
\end{figure}

We show current CMB bounds in the $(\ghlightCMB,\ghheavyCMB)$ plane in
\figref{gstarcmb}, along with four example models.
Figure~\ref{fig:gstar}'s Model C, with all degrees of freedom light
compared to $\ThBBN$, is now differentiated into many models,
depending on how many of the degrees of freedom are light compared to
$\ThCMB$. These define a line in \figref{gstarcmb}, with endpoints
given by Models C$^\prime$ and C$^{\prime \prime}$.  We find that if
these models are allowed by the BBN constraints, they are also allowed
by the current CMB constraint.  Similarly, \figref{gstar}'s Model D is
differentiated into Models D$^\prime$ and D$^{\prime \prime}$ in
\figref{gstarcmb}, but again we find that in such models, if the BBN
constraint is satisfied, the current CMB constraint is also.

In these models, then, the current CMB constraint is never as
stringent as the BBN constraint.  This may change with the
strengthening of the CMB constraint in the future.  In addition, it is
interesting that in the presence of hidden sectors, there may be many
particles with mass between $\ThBBN$ and $\ThCMB$, and BBN and the CMB
yield independent information about the hidden sector mass spectrum.
This contrasts with the case of the SM, where CMB constraints may be
viewed as consistency checks on BBN results.

\section{A Concrete Model}
\label{sec:concrete}

In \secref{BBN}, we found that hidden sectors may contain many degrees
of freedom without violating cosmological bounds.  With this as
motivation, in the following section, we will examine freezeout in
hidden sectors whose temperatures may differ from the observable
sector's. To prepare the way, in this section we first specify a model
and mass spectrum for the hidden sector.

The WIMPless models of Ref.~\cite{Feng:2008ya} are supersymmetric,
with supersymmetry breaking mediated by gauge interactions.  A
worthwhile endeavor would be to build a complete GMSB WIMPless model.
In this study, however, we instead work with a ``GMSB-inspired''
model.  We do this for two reasons.  First, our primary aim here is to
clarify cosmological issues.  Building a complete GMSB model with a
WIMPless hidden sector, complete with $\mu$ term solution and other
particle physics details, would take us too far afield, especially
since, as we will see, the relic density is independent of many of
these details.  Second, as noted in \secref{intro}, the essential
virtue of WIMPless models, that they have the naturally correct
thermal relic density, relies solely on the fact that masses in the
hidden sector are proportional to gauge couplings squared.  This is
true for GMSB, of course, but is also valid in many other contexts,
such as anomaly-mediated supersymmetry breaking models,\footnote{We
thank Raman Sundrum for making this point.} lessening the motivation
for detailed consideration of any one GMSB model.

The model we consider has an MSSM-like hidden sector, with
$\text{SU(3)} \times \text{SU(2)} \times \text{U(1)}$ gauge
interactions.  In contrast to the MSSM, however, we assume that all
Yukawa couplings are ${\cal O}(1)$.  We refer to this as a flavor-free
version of the MSSM.  For simplicity, we will focus on a 1-generation
flavor-free version of the MSSM, but, as all Yukawa couplings are
large, we will use 3rd generation nomenclature for this one generation
of hidden matter fermions and sfermions.

All hidden superpartners are assumed to have mass $\sim m_X$, the
superpartner mass scale.  In addition, since $m_X$ also sets the scale
for the Higgs boson parameters, the hidden Higgs bosons $h^h$, $H^h$,
$A^h$, and $H^{h\pm}$ also have this mass, as do the hidden weak gauge
bosons $W^h$ and $Z^h$.  Given ${\cal O}(1)$ Yukawa couplings, the
hidden fermions $\tau^h$, $b^h$, and $t^h$ are therefore also at the
superpartner mass scale.  The light, effectively massless particles
are the hidden photon $\gamma^h$, gluon $g^h$ and neutrino $\nu^h$.
As is typical of GMSB, the gravitino is also light, and its relic
density from late decays to the gravitino is negligible.  We further
assume low or moderate reheating temperatures so that the abundance of
gravitinos produced during reheating is also insignificant.

WIMPless dark matter requires that one particle at the $m_X$ mass
scale be stable. We assume that the lightest $m_X$ scale particle is
the hidden right-handed stau $X \equiv \tilde{\tau}^h_R$.  This is the
lightest superpartner in much of GMSB parameter space, and it is also
reasonable to assume that it is lighter than the $\tau^h$, given
${\cal O}(1)$ Yukawa couplings, just as the stop may be lighter than
the top in the MSSM.  The stau's stability is then guaranteed by
hidden sector electric charge conservation.  One might worry about the
viability of a charged dark matter candidate.  Efficient Compton
scattering of hidden staus on hidden photons $\stau^{h \pm}_R \gamma^h
\to \stau^{h \pm}_R \gamma^h$ would lead to dissipative energy loss,
making the dark matter particles behave more like baryons.  However,
this scattering cross section is $\sigma \sim g^{h\, 4}/m_X^2$, which,
given \eqref{scaling}, is as weak as the SM weak interactions.  Hidden
sector charged dark matter is therefore consistent with the known
properties of dark matter from structure formation.

Given the few light states, there are only a few dark matter
annihilation channels:
\begin{equation}
\stau_R^{h\, +} \stau_R^{h\, -} \to \nu^h \bar{\nu}^h ,
\gamma^h \gamma^h, \gamma^h Z^h \ .
\end{equation}
The neutrino channel is mediated by an $s$-channel $Z^h$.  The gauge
boson final states are produced through processes with $\stau_R^h$ in
$t$- and $u$-channels and 4-point $\stau \stau \gamma \gamma$ and
$\stau \stau \gamma Z$ vertices.  The relic density is therefore
dependent on only a few of the many supersymmetric parameters, which
we will now discuss.  Note that we assume no left-right stau mixing.

All three gauge couplings enter the analysis.  The SU(2) and U(1)
gauge couplings enter the annihilation cross sections.  The SU(3)
gauge coupling must also be specified, because it determines the
number of relativistic degrees of freedom at BBN.  For simplicity, for
the hidden gauge couplings, we adopt assumptions motivated by
unification.  Gauge coupling unification implies that the hidden weak
mixing angle is $\tan \theta^h_W = \sqrt{3/5}$ at the GUT scale.  In
principle, this is modified by renormalization group (RG) evolution.
The RG equations for the hidden sector MSSM gauge couplings are
\begin{eqnarray}
\frac{d g^h_a}{d \ln (Q/M )}=\frac{1}{16\pi^2} b^h_a\, g^{h\, 3}_a \ ,
\end{eqnarray}
where $b^h_a=(13/5,-3,-7)$ for the 1-generation case, and
$b^h_a=(33/5,1,-3)$ for the 3-generation case.  As we will see in
\secref{relic}, however, WIMPless models are valid in the ranges $\kev
\alt m_X < \tev$ and $10^{-5} \alt g_X(m_X) < 1$, depending on the
reheating temperature ratio $\xiRH$.  For most of this range, then,
$g_X (m_X) \alt 0.1$, and the fractional change in the gauge couplings
from RG evolution is $| \Delta g^h_a/g^h_a | \approx |b^h_a| (g^{h\,
2}_a/ 16 \pi^2) \ln (M_{\text{mess}} / m_X) \alt 0.01$; that is, the
gauge couplings evolve very little.  Given this, we will assume $\tan
\theta^h_W = \sqrt{3/5}$ at $m_X$. For $m_X \agt 100~\gev$, this is
not the prediction of unification, and our assumption therefore
implicitly assumes some more complicated structure.

The unification assumption also typically implies that the hidden
confinement scale
\begin{equation}
   \Lambda^h_{\text{QCD}} \sim
m_X\, e^{\frac{8\pi^2}{b^h_3 g^{h\, 2}_3(m_X)}}
\end{equation}
is far below $\ThBBN$.  For example, for the 1-generation case, taking
$m_X \sim 100~\gev$ and $g_3(m_X)=0.5$, we find $\Lambda^h_{\text{QCD}}
\sim 10^{-9}~\text{eV}$.  We therefore assume $\Lambda^h_{\text{QCD}}
< \ThBBN$, and so the hidden gluons contribute to $g^h_*$ at
BBN. Again, this does not follow from unification for the largest
$m_X$.

We will identify $g_X$ with the hidden SU(2) gauge coupling $g^h$.  As
for the hidden MSSM mass spectrum, we assume that all massive hidden
particles have masses in the range $m_X \leq m_i \leq 2 m_X$, and that
all relativistic degrees of freedom are always at the same
temperature.  The effective relativistic degrees of freedom in the
hidden sector are then
\begin{equation}
\label{gstar_model}
   g^h_\ast (T^h) = g^h_{\ast S} (T^h) = \left\{ \begin{array}{l}
   116.25\, (228.75) \, , \ T^h \geq 2 m_X \\
   19.75\, (23.25) \, , \quad \ T^h \leq m_X \ ,
   \end{array} \right.
\end{equation}
for the 1-generation (3-generation) case, and we linearly interpolate
between these two values for $T^h$ between $m_X$ and $2 m_X$. The
exact spectrum in this range is largely irrelevant for the relic
abundance calculation, since we consider only dark matter that is
non-relativistic at freezeout.  Given these assumptions, the thermal
relic density is completely determined by only four parameters:
\begin{equation}
m_X , \ g_X , \ m_Z , \ \xiRH \equiv \frac{\ThRH}{\TRH} \ .
\end{equation}

\begin{figure}[tb]
\begin{center}
\includegraphics*[width=12cm,clip=]{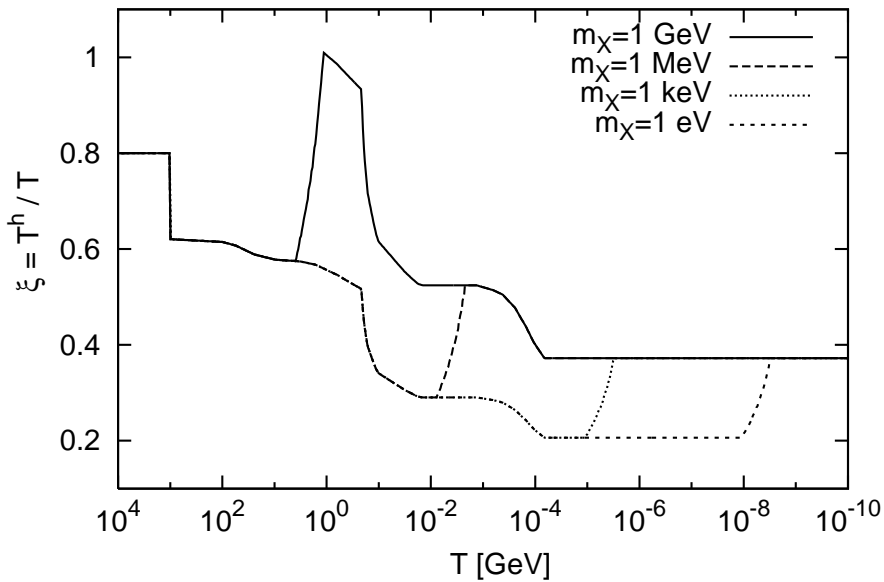}
\end{center}
\vspace*{-.29in}
\caption{Evolution of $\xi = T^h/T$, the ratio of hidden to
observable temperatures, for $m_X = 1~\gev$, 1 MeV, 1 keV and 1 eV,
assuming $\TRH=50~\tev$, $\xiRH=0.8$, and a hidden sector that is
the 1-generation flavor-free version of the MSSM.  For other
$\xiRH$, these curves are simply re-scaled by $\xiRH / 0.8$. All
supersymmetric particles in the observable sector are assumed to
have mass around 1 TeV.} \label{fig:xi}
\end{figure}

In \figref{xi} we plot the relative evolution of the observable and
hidden photon temperatures for the cases $m_X = 1~\gev$, 1 MeV, 1
keV and 1 eV. To make this plot, we must determine $T^h$ as a
function of $T$ for a given $m_X$.  To get the hidden sector
temperature $T^h$ at a given observable sector temperature $T$, the
value of $g^h_{\ast S} (T^h)$ is required, which is in fact
unattainable unless one knows $T^h$. Our approach is the following:
we first solve for $g^h_{\ast S} (T^h)\, T^{h\, 3}$ from
\eqref{entropy} with $\TBBN$ and $\ThBBN$ replaced by $T$ and $T^h$,
respectively. We then determine $T^h$ by dividing this quantity by
$g^h_{\ast S} (T^h)$ as determined by \eqref{gstar_model} in the
three trial intervals $T^h \le m_X$, $m_X < T^h < 2 m_X$, and $2 m_X
< T^h$, and choose the value of $T^h$ that yields a self-consistent
solution.

The general feature of \figref{xi} is that as $T$ decreases from
$\TRH$, $\xi = T^h/T$ drops when observable degrees of freedom become
non-relativistic and rises when hidden degrees of freedom become
non-relativistic.  Rapid drops occur at $T \sim \tev$ when the
observable superpartners become non-relativistic and $\xi \propto
\sqrt[3]{g_{\ast S} (T)}$, and at $T \sim 250~\mev$, where the QCD
transition occurs.  The rise occurs when the hidden $m_X$ mass
particles become non-relativistic at $T \sim 4 m_X - 10 m_X$.  At $T
\sim m_X - 5 m_X$, where the exact value depends on $m_X$, the hidden
sector phase transition concludes, and $\xi$ begins to track
$\sqrt[3]{g_{\ast S} (T)}$ again.

\section{Thermal Relic Densities}
\label{sec:relic}

We are now ready to determine thermal relic densities in the case of
observable and hidden sectors with unequal temperatures.  For the
hidden sector, we assume the 1-generation flavor-free version of the
MSSM described in \secref{concrete}.  The temperature of the hidden
sector thermal bath at any instant can be read off from \figref{xi}.

The number density of dark matter particles $n$ is determined by the
competition between the expansion rate of the universe and the dark
matter annihilation rate through the Boltzmann equation
\begin{equation}
\frac{dn}{dt} = -3 H n - \langle \sigma_A v \rangle (T^h)
\left[ n^2 - n_{\text{eq}}^2 (T^h) \right] \ .
\end{equation}
The Hubble parameter depends on the effective relativistic degrees of
freedom in both the observable and hidden sectors and is
\begin{equation}
H (T) = \left[ \frac{4 \pi^3 G_N}{45} g_*^{\text{tot}}(T) \, T^4
\right]^{\frac{1}{2}} \ ,
\end{equation}
where
\begin{equation}
g_*^{\text{tot}} (T) =
g_*(T) + g^h_* (T^h) \left( \frac{T^h}{T} \right)^4 \ .
\end{equation}
In contrast to $H$, $\langle \sigma_A v \rangle$, the
thermally-averaged product of the total annihilation cross section and
the M{\o}ller velocity, and $n_{\text{eq}}$, the equilibrium number
density, are determined by the hidden temperature alone.

To solve the Boltzmann equation, it is standard to change to
dimensionless variables through
\begin{equation}
t \to x \equiv \frac{m_X}{T} \qquad \qquad
n \to Y \equiv \frac{n}{s} \ .
\label{observablevariables}
\end{equation}
These new variables are natural from the observable sector viewpoint,
as they use the observable sector's temperature $T$ as the ``clock,''
and the observable sector's entropy density $s = \frac{2 \pi^2}{45}\,
g_{*S}\, T^3$ as the fiducial quantity characterizing the universe's
expansion.  Note, however, that in the present context there is also
another useful change of variables,
\begin{equation}
t \to x^h \equiv \frac{m_X}{T^h} \qquad \qquad
n \to Y^h \equiv \frac{n}{s^h} \ ,
\end{equation}
that is far more natural from the hidden sector point of view.  We
will return to this below.

In terms of the observable sector variables of
\eqref{observablevariables}, the Boltzmann equation becomes
\begin{equation}
\label{Boltzmann_eq_Y}
  \frac{dY}{dx} = - \lambda \left[ Y^2 - Y^2_{\text{eq}} (x^h) \right]\, ,
\end{equation}
with the expansion and annihilation effects encapsulated in the
quantity
\begin{equation}
\label{eq:lambda}
\lambda = \sqrt{\frac{\pi}{45\, G_N}}\,
\frac{g_{*S} (T)}{\sqrt{g_*^{\text{tot}} (T)}}\,
\left(1 + \frac{1}{3}\, \frac{T}{g_{*S} (T)}\,
\frac{d g_{*S} (T)}{d T} \right)\,
\frac{m_X}{x^2}\, \langle \sigma_A v \rangle (T^h)\ .
\end{equation}
We will solve this equation in the non-relativistic regime, where
$x^h \agt 3$. The equilibrium value of the number of dark matter
particles in a comoving volume is
\begin{equation}
Y_{\text{eq}}(x^h)
\equiv \frac{n_{\text{eq}}(T^h)}{s(T)}
= \frac{1}{s(T)}\frac{g}{2\pi^2}\int^{\infty}_{m_X}
\frac{\sqrt{E^2-m^2_X}}{e^{E/T^h}-1} E dE
\approx\frac{45 x^{h\, 2}}{4 \pi^4\, g_{*S} (T)} g\, K_2 (x^h)\xi^3\
, \label{yeq}
\end{equation}
where $g$ denotes the number of internal degrees of freedom of the
dark matter particle, and the last approximation of \eqref{yeq} is
valid for $x^h \agt 3$. Following Ref.~\cite{Gondolo:1990dk}, the
thermally-averaged cross section times velocity is determined using
\begin{equation}
\label{sigmav}
\langle \sigma_A v \rangle (T^h) =
\frac{1}{8 m_X^4 T^h K^2_2 \! \left( \frac{m_X}{T^h} \right)}
\int^{\infty}_{4 m_X^2} ds\, \sigma_A(s) \sqrt{s} (s-4 m_X^2)
   K_1 \! \left(\frac{\sqrt{s}}{T^h}\right) .
\end{equation}
In \eqsref{yeq}{sigmav}, $K_i$ are modified Bessel functions of order
$i$. Equation~(\ref{sigmav}) was derived for particles in thermal
equilibrium obeying Maxwell-Boltzmann statistics, but it is applicable
to other statistics, provided $T^h \alt m_X/3$.

We determine the $\stau^h_R$ total annihilation cross section
$\sigma_A (s)$ with the help of the CalcHEP
package~\cite{Pukhov:1999gg,Pukhov:2004ca}. Expanding around $\epsilon
\equiv (s - 4 m^2_X) /4 m^2_X$ (see Ref.~\cite{Gondolo:1990dk}), we
find
\begin{equation}
\label{partial_wave}
   \langle \sigma_A v \rangle (T^h) = \frac{g^4_X}{m^2_X} \left[ a_0 +
   a_1\, \left(\frac{m_X}{T^h} \right)^{-1} + a_2\,
   \left(\frac{m_X}{T^h} \right)^{-2} + \ldots \right] \, ,
\end{equation}
where
\begin{eqnarray}
   a_0 &=& \left[\frac{1}{8 \pi} + \frac{1}{4 \pi}\,
   \left(1 - \frac{m^2_{Z^h}}{4 m^2_X} \right)\,
   \tan^2 \theta^h_W \right]\,
   \sin^4 \theta^h_W \\
   a_1 &=& \frac{3}{2}\, \left[- \frac{1}{6 \pi} - \frac{1}{3 \pi}\,
   \tan^2 \theta^h_W + \frac{1}{12 \pi}\, \frac{1}{\cos^4 \theta^h_W\,
   \left[\left(-4 + \frac{m^2_{Z^h}}{m^2_X} \right)^2 + \frac{m^2_{Z^h}
   \Gamma^2_{Z^h}}{m^4_X} \right]} \right]\, \sin^4 \theta^h_W\ ,
\end{eqnarray}
and $\Gamma_{Z^h}$ is the $Z^h$ boson decay width.  In keeping with
our non-relativistic approximation, we neglect the $a_2$ and higher
order terms.  For $m_{Z^h} \approx 2 m_X$, $\stau^{h\, +}_R \stau^{h\,
-}_R \to \nu^h \bar{\nu}^h$ is greatly enhanced by the $s$-channel
resonance.  Off this resonance, however, the neutrino channel is
typically sub-dominant, because it is $P$-wave suppressed, as can be
seen above.  In our numerical study, we fix $m_{Z^h} = 1.5\, m_X$.
The $Z^h$ boson remains in equilibrium with the hidden thermal bath
through the decay and inverse decay processes $Z^h \leftrightarrow
\nu^h \bar{\nu}^h$.

\begin{figure}[tb]
\begin{center}
\includegraphics*[width=12cm,clip=]{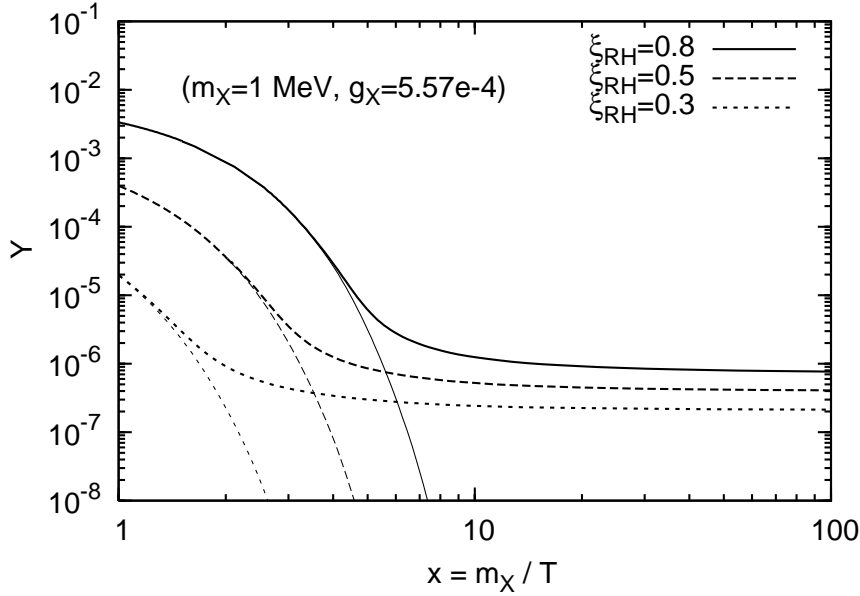}
\end{center}
\vspace*{-.29in}
\caption{$Y(x)$, the $\stau^h_R$ number density per comoving volume,
as a function of $x \equiv m_X/T$, for $m_X = 1~\mev$, $g_X = 5.57
\cdot 10^{-4}$, and various ratios of hidden sector to visible sector
reheating temperature $\xiRH = 0.8$, 0.5 and 0.3 (from top to
bottom).}
\label{fig:freezeout_xi}
\end{figure}

We now present several results for $Y(x)$.  Because of the
``stiffness'' of the Boltzmann equation of \eqref{Boltzmann_eq_Y}, we
adopt the implicit trapezoidal method with adaptive step size,
developed by the authors of the DarkSUSY
package~\cite{Gondolo:2004sc}, to solve for $Y(x)$ numerically.  We
first examine the $\xiRH$ dependence of the thermal relic density.  In
\figref{freezeout_xi}, we plot $Y(x)$ for various values of $\xiRH$,
$m_X = 1~\mev$, and $g_X = 5.57 \cdot 10^{-4}$.  The latter two
parameters are naturally accommodated in the WIMPless GMSB framework
and yield $\Omega_X h^2$ in the required range, as we will see below.
As $\xiRH$ decreases, we see two effects.  First, for a given $x$, the
hidden sector is colder, and so $Y_{\text{eq}}(x)$ is smaller.
Second, since freezeout occurs when $H \sim \Gamma = n \langle
\sigma_A v \rangle$, when $\xiRH$ decreases and $n$ drops, freezeout
occurs earlier.  As can be seen from \figref{freezeout_xi}, these
effects act against each other, and the thermal relic density is
fairly insensitive to $\xiRH$.

These results may be understood more precisely using standard analytic
approximation formulas for thermal relic freezeout.  The equilibrium
number density is
\begin{equation}
Y_{\text{eq}}=0.145(g/g_{*S})x^{3/2}\xi^{3/2}e^{-x/\xi} \ ,
\label{yeqapprox}
\end{equation}
where $\xi= T_h/T$, as defined above. Following the procedure given in
Ref.~\cite{Gondolo:1990dk} and assuming $S$-wave annihilation, an
approximate expression for $x$ at freezeout is
\begin{eqnarray}
x_f &\approx& \xi\ln \left[ 0.038 \, \mplanck \, m_X \, \sigma_0
  (g/\sqrt{g^{\text{tot}}_*}) \, \xi^{3/2}\, \delta(\delta+2) \right]
  \nonumber \\
&& -\frac{1}{2} \xi \ln \left\{ \xi \ln \left[ 0.038 \,
\mplanck \, m_X \, \sigma_0 (g/\sqrt{g^{\text{tot}}_*}) \,
\xi^{3/2} \, \delta(\delta+2) \right] \right\} \ ,
\label{xf}
\end{eqnarray}
where $\sigma_0 = a_0 g^4_X / m^2_X$ and the parameter $\delta$ is
tuned to make these analytical results fit the numerical results.
\footnote{For $\xiRH=0.3 \ (0.8)$, we find that choosing $\delta = 0.2
\ (0.5)$ yields agreement typically better than 3\%. For larger $m_X$
or higher $\xiRH$, the final relic density is less sensitive to
$\delta$.}

For $S$-wave annihilation, the final relic abundance is approximately
\begin{equation}
\label{Y}
Y_0 \approx \frac{3.79 \, x_f}
{\left( g_{*S}/\sqrt{g^{\text{tot}}_*} \right)
  \mplanck \, m_X \, \sigma_0} \ ,
  \label{Y0}
\end{equation}
which depends linearly on $\xi$ through $x_f$. This linear dependence
is verified by the numerical results presented in
\figref{freezeout_xi}. The relative values of $Y_0$ obtained from
\eqref{Y0} are $3.6:1.9:1$ for $\xiRH = 0.8$, 0.5, 0.3, respectively,
which almost exactly match the results from our numerical computation,
which yields $3.5:1.9:1$.

\begin{figure}[tb]
\begin{center}
\includegraphics*[width=12cm,clip=]{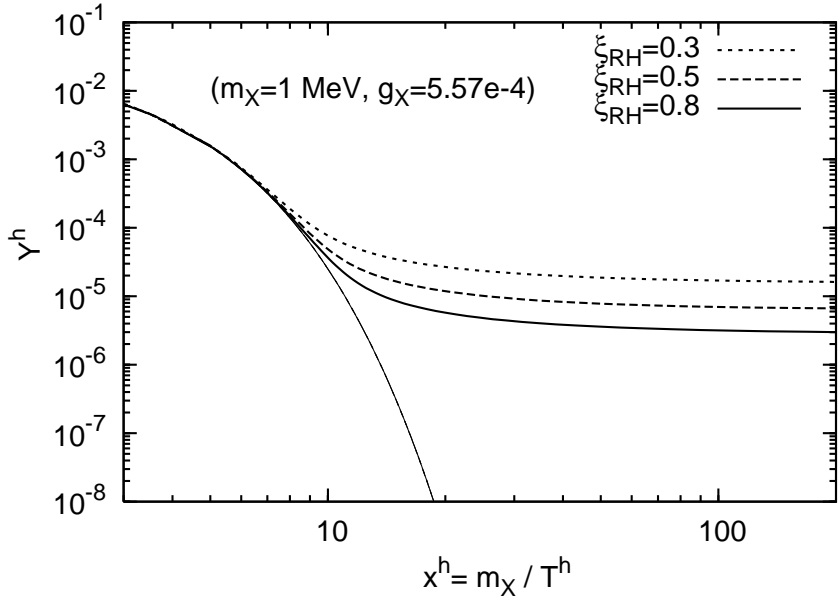}
\end{center}
\vspace*{-.29in}
\caption{As in \figref{freezeout_xi}, but now plotted in terms of the
  hidden sector parameters $Y^h(x^h)$.}
\label{fig:freezeout_xi_hidden}
\end{figure}

The dependence on $\xiRH$ may also be understood from the hidden
sector viewpoint.  In \figref{freezeout_xi_hidden}, we consider the
same model parameters as plotted in \figref{freezeout_xi}, but we now
present $Y^h(x^h)$.  {}From a hidden observer's point of view, the
dependence on $\xiRH$ arises solely because $\xiRH$ determines how hot
the observable sector is, and this impacts the expansion rate $H$.
For a fixed $x^h$, lower $\xiRH$ implies higher $T$, faster expansion,
and earlier freezeout.  This is seen in \figref{freezeout_xi_hidden}.
At first sight, the result that lower $\xiRH$ implies larger $Y^h$ may
seem to contradict the results obtained earlier, where lower $\xiRH$
implies lower relic density.  Note, however, that to convert $Y^h$ to
a relic density, one must multiply $Y^h$ by the current $s^h$, which,
unlike $s$, is not independent of $\xiRH$.  For lower $\xiRH$, the
current hidden sector temperature is lower, and the current $s^h$ is
also lower.  This effect makes the relic density smaller for lower
$\xiRH$, in accord with our previous results derived from the
observable sector viewpoint.

\begin{figure}[tb]
\begin{center}
\includegraphics*[width=12cm,clip=]{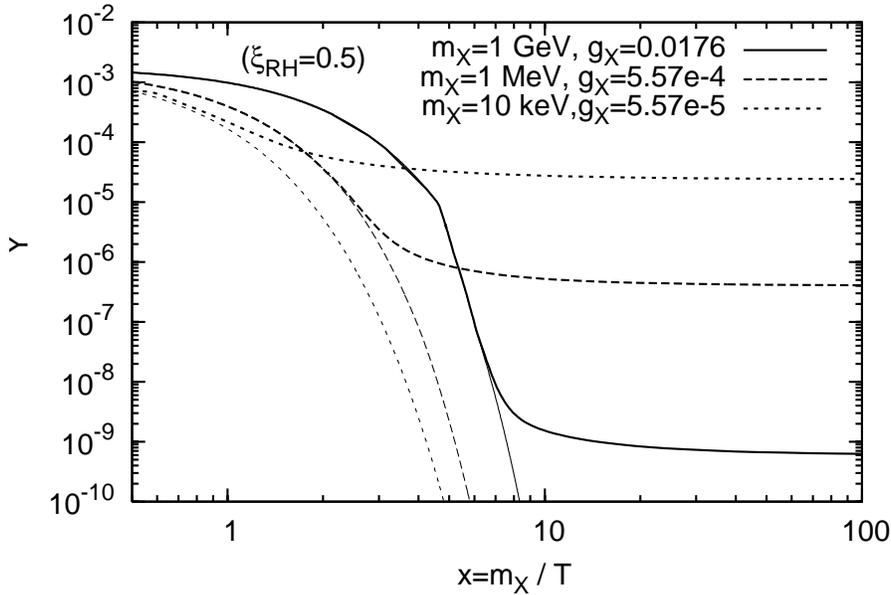}
\end{center}
\vspace*{-.29in}
\caption{$Y(x)$, the $\stau^h_R$ number density per comoving volume,
as a function of $x \equiv m_X/T$, for fixed $\xiRH = 0.5$ and
$(m_X, g_X) = (1~\gev, 1.76 \cdot 10^{-2})$, $(1~\mev, 5.57 \cdot
10^{-4})$, and $(10~\kev, 5.57 \cdot 10^{-5})$. All cases are
chosen to have the same ratio $m_X / g^2_X$.}
\label{fig:freezeout_mx}
\end{figure}

We now examine the dependence of the thermal freezeout process on
$m_X$ and $g_X$.  In \figref{freezeout_mx}, we plot $Y(x)$ for fixed
$\xiRH$, but for a wide range of $(m_X, g_X)$ with $m_X \propto
g_X^2$, the scaling relation of WIMPless scenarios.  For low $m_X$,
freezeout occurs earlier, as expected given \eqref{xf}.  In our
analysis, we have assumed that freezeout occurs in the
non-relativistic regime, where $x^h \agt 3$.  Converting $x^h$ to $x$
using \figref{xi}, we see from \figref{freezeout_mx} that our analysis
is self-consistent even for $m_X$ as low as 10 keV.

\begin{figure}[tb]
\begin{center}
\includegraphics*[width=12cm,clip=]{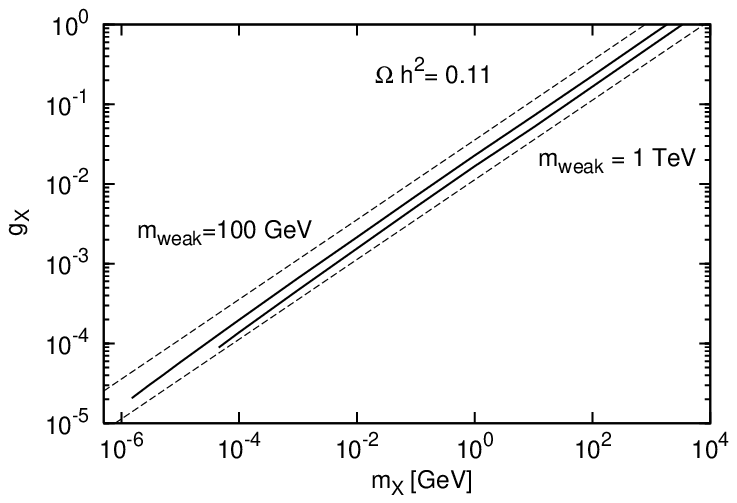}
\end{center}
\vspace*{-.29in}
\caption{Contours of $\Omega_X h^2 = 0.11$ in the $(m_X,g_X)$ plane
for $\xiRH=0.8$ (upper solid) and 0.3 (lower solid), where the hidden
sector is a 1-generation flavor-free version of the MSSM,
corresponding to Model A (for $m_X \geq 1~\mev$) or Model C (for $m_X
< 1~\mev$).  The minimum $m_X$ consistent with the requirement of
non-relativistic freezeout depends strongly on $\xiRH$ (see text).
Also plotted are lines of $\mweak \equiv (m_X/g^2_X)g'^2 = 100~\gev$
(upper dashed) and 1 TeV (lower dashed).}
\label{fig:WMAPgxmx}
\end{figure}

In \figref{WMAPgxmx} we present contours of constant $\Omega_X h^2$ in
the $(m_X, g_X)$ plane.  These contours result from our numerical
analysis; although we have confined our previous plots to $x \alt
100$, the results of \figref{WMAPgxmx} use the relic density as
determined at $x \sim 1000$, where it has truly stabilized.

Figure~\ref{fig:WMAPgxmx} summarizes all of our work so far, and there
are several noteworthy features.  First, we see that these contours
essentially follow the scaling relation $m_X \propto g_X^2$.  If, for
example, $m_X$ is lowered by a factor of 4 and $g_X$ is lowered by a
factor of 2, the weakened interaction implies earlier freezeout and
larger $Y_0$, but this compensates the lower $m_X$ to keep the relic
density $\Omega_X h^2 \propto m_X Y_0$ approximately invariant.
Alternatively, using the analytic results presented above, this
scaling relation keeps $\sigma_A$ invariant, and, given \eqref{Y0},
this also keeps $\Omega_X h^2$ invariant.  Of course, this is as
expected from the general arguments given in \secref{intro}, but
\figref{WMAPgxmx} shows that these arguments, based essentially on
dimensional analysis, are in fact supported by a detailed calculation
in the context of a concrete model with dark matter masses varying
greatly and far beyond the conventional WIMP range.

Second, for comparison, \figref{WMAPgxmx} also includes two dashed
contours to indicate the corresponding observable sector weak mass
scale $\mweak$.  The definition of $\mweak$ depends on how one chooses
to characterize the natural weak scale in supersymmetric theories (the
$\mu$ parameter, the Bino mass, etc.), and these depend on several
parameters, such as the number of messengers and the messenger scale.
We simply choose $\mweak \equiv (m_X/g^2_X)g'^2$, where $g'$ is the
U(1) gauge coupling of the MSSM, and plot contours for $\mweak =
100~\gev$ and 1 TeV. The fact that there exist values in the $(m_X,
g_X)$ plane that give the correct $\Omega_X h^2$ is, of course, not
surprising.  However, the fact that the two solid curves lie between
the two dashed curves means that the preferred values of $(m_X, g_X)$
are those that correspond to models in which the observable sector's
superpartners are at the weak scale.

We also see from \figref{WMAPgxmx} that, for a fixed $m_X$, lower
$\xiRH$ requires lower $g_X$ to obtain the correct relic density.
This is because, at a given observable temperature $T$, colder hidden
sectors imply lower equilibrium number densities.  To keep the relic
density constant, $g_X$ must be lower so that freezeout happens
earlier.  This is consistent with the results shown in
\figref{freezeout_xi}.

\begin{figure}[tb]
\begin{center}
\includegraphics*[width=12cm,clip=]{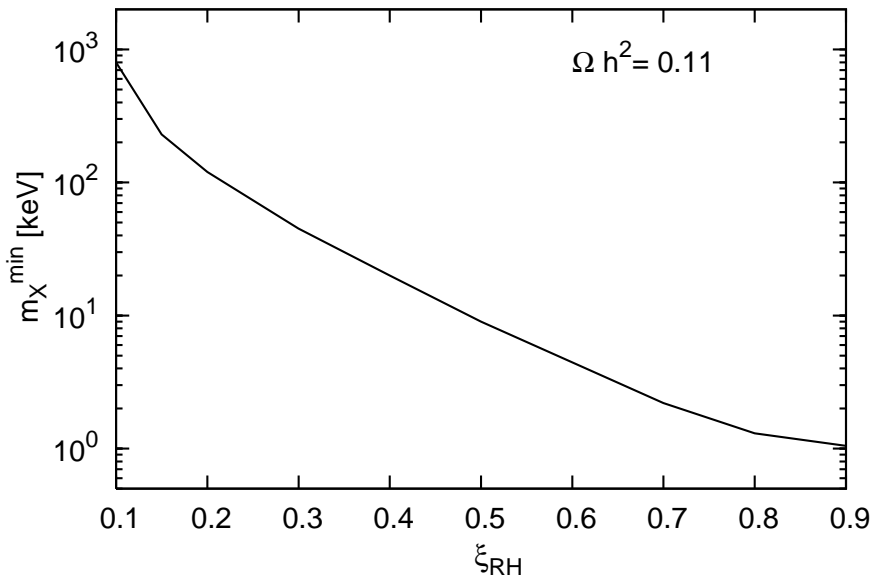}
\end{center}
\vspace*{-.29in}
\caption{Lower bound on the dark matter particle mass $m_X$ for
different $\xiRH$ in the WIMPless scenario. As in \figref{WMAPgxmx},
the hidden sector is a 1-generation flavor-free version of the MSSM,
corresponding to Model A (for $m_X \geq 1~\mev$) or Model C (for $m_X
< 1~\mev$). }
\label{fig:mxin}
\end{figure}

Finally, note that the $\xiRH = 0.3$ contour ends at $m_X\simeq
45~\kev$.  This is because the criterion that dark matter freezeout
while non-relativistic, that is, $x^h_f \agt 3$, is violated below
this mass. In \figref{mxin}, we plot, as a function of $\xiRH$, the
lower mass limit resulting from requiring that freezeout occurs with
$x^h_f \agt 3$ and yields $\Omega h^2=0.11$. We can see that the lower
mass limit goes up with smaller $\xiRH$. This is because for colder
hidden sectors, freezeout occurs earlier, and to satisfy $x^h_f \agt
3$, we need larger $m_X$. This exercise establishes that the WIMPless
dark matter framework may be valid at least down to dark matter masses
of $m_X \sim \kev$.  Below this mass, freezeout is not
non-relativistic, and our analysis breaks down.  It would be
interesting to perform a more precise analysis without assuming
non-relativistic freezeout, but this is beyond the scope of our study.

\section{Implications of Connectors}
\label{sec:adding}

Throughout this work, we have ignored the possible existence of
connector particles $Y$ with both SM and hidden gauge charges.  If
these are added, do they significantly affect the conclusions derived
above?

There are two significant roles that connector particles can have.
First, they likely establish thermal connectivity between the
observable and hidden sectors when the temperatures are above the
connector mass scale $m_Y$.  At temperatures below $m_Y$, however, the
hidden-SM interactions become weak, and the observable and hidden
sectors decouple.  As shown in \figsref{gstar}{gstarcmb}, the BBN and
CMB constraints actually allow, for example, a 1-generation
flavor-free version of the MSSM with $m_X \alt \ThBBN$, even when
$\xiRH = 1$.  In this case, then, there is no problem if the two
sectors are in thermal contact right after reheating. For other hidden
sectors, it may be possible that the existence of connectors enforces
$\xiRH = 1$, or alternatively, the assumption of $\xiRH \ne 1$ implies
an upper bound on the reheating temperatures.  One must check on a
case by case basis whether connector particles upset the premises of
our calculations.

A second affect is that connector particles open new avenues for dark
matter annihilation.  As an example, we assume that fermionic
connector particles $Y$ couple through Yukawa couplings $\lambda X
\bar{Y} f$, where $f$ is a SM fermion.  We expect $m_Y \sim \max (
m_X, \mweak)$~\cite{Feng:2008ya}, and also require $m_Y > m_X$ to
prevent the decay $X \to Y f$.

Such connector interactions induce annihilations $XX \to \bar{f} f$
through processes with $Y$ particles in the $t$-channel.  There are
two cases.  If $m_X \alt \mweak$, $m_Y \sim \mweak$, and
\begin{equation}
\langle \sigma v \rangle_{\bar{f} f}
\sim \frac{\lambda^4 m_X^2}{m_Y^4}
\alt \frac{\lambda^4}{m_Y^2} \sim \frac{\lambda^4}{\mweak^2} \ .
\end{equation}
On the other hand, if $m_X \agt \mweak$, $m_Y \sim m_X$, and
\begin{equation}
\langle \sigma v \rangle_{\bar{f} f}
\sim \frac{\lambda^4}{m_X^2}
\alt \frac{\lambda^4}{\mweak^2} \ .
\end{equation}
In either case, the annihilation cross section to SM particles is
small relative to the completely hidden annihilation cross sections
$\gweak^4/\mweak^2$, provided $\lambda \alt \gweak \simeq 0.65$. This
is a rather weak criterion, and so typically, annihilation to SM
particles is sub-dominant, and the analysis given above holds even in
the presence of connector particles.

\section{Conclusions}
\label{sec:conclusions}

In the paper, we have studied the possibility that dark matter is
hidden, that is, has no SM gauge interactions.  Hidden sectors appear
in many frameworks for physics beyond the SM, and hidden dark matter
is perfectly viable, given all observations to date.  We are
particularly motivated by the recent proposal of WIMPless dark matter.
As with WIMPs, WIMPless dark matter has the key virtue that its
thermal relic density is in the right range to be cold dark matter.
In contrast to WIMPs, however, WIMPless dark matter may have masses
and couplings that differ drastically from WIMPs.

Although hidden sectors interact very weakly with the SM, hidden
particles contribute to the energy density of the Universe.  This
impacts the expansion rate of the Universe, and so bounds from BBN and
the CMB constrain hidden sectors.  As is well known, current
constraints completely exclude a hidden sector that is an exact copy
of the MSSM.  We find, however, that these constraints are rather
brittle and easily avoided if the hidden sector is just slightly
colder than the observable sector or if the mass scale of hidden
superpartners differs from the MSSM.  Several viable examples are
discussed in \secref{BBN}, and \figsref{gstar}{gstarcmb} give
model-independent bounds that can be used to determine the viability
of other models.

In \secref{concrete}, we then defined a concrete model, a 1-generation
flavor-free version of the MSSM.  This model has features generic to
GMSB models and provides a WIMPless dark matter candidate, the hidden
stau $\stau_R^h$.  The $\stau_R^h$ relic density depends sensitively
on only a small number of parameters.  This relic density was examined
in \secref{relic} by solving the Boltzmann equation numerically, and
understanding these results through well-known analytic
approximations, generalized to hidden sectors with different
temperatures.  These results confirm that in WIMPless models with $m_X
\propto g_X^2$, the relic density is naturally in the right range for
dark matter masses $\kev \alt m_X \alt \tev$, greatly extending the
conventional WIMP mass range.  WIMPless dark matter therefore provides
a class of dark matter candidates that shares the key relic density
virtue of WIMPs.  This generalization may have interesting
applications, given the diverse and interesting phenomenology and
observational anomalies already present at the keV, MeV, GeV, and TeV
mass scales.

We have determined the allowed range of $m_X$ by requiring that this
dark matter freezes out with the right relic density while
non-relativistic.  Of course, this range may be further constrained by
other considerations.  In particular, the lightest WIMPless
candidates, with $m_X \sim \kev$, may have interesting implications
for structure formation.  After freeze out, they can still couple to
the hidden sector thermal bath through elastic scattering processes
$\stau_R^h \gamma^h \to \stau_R^h \gamma^h$ and $\stau_R^h \nu^h \to
\stau_R^h \nu^h$.  After their kinetic decoupling at temperature
$T_{\text{kd}}$, they then stream freely between over- and under-dense
regions.  Free-streaming of thermal relics causes damping in density
perturbations~\cite{Bertschinger:2006nq} at a comoving scale
$\lambda_{\text{FS}} \propto m^{-1/2}_X\, T_{\text{kd}}^{-1/2}$; in
contrast to WIMPs~\cite{Loeb:2005pm,Profumo:2006bv}, this is the
dominant effect for MeV or lighter dark matter~\cite{Hooper:2007tu}.
If this scale falls in the range $(1-80)\, h^{-1}$ Mpc, the linear
matter power spectrum from high-resolution Lyman-$\alpha$ forest data
can be used to place a lower bound on the mass of the dark matter
particle. Current lower mass limits on warm dark matter of $0.55 -
{\cal O}(1)~\kev$ from Lyman-$\alpha$
data~\cite{Narayanan:2000tp,Viel:2005qj,Abazajian:2005xn,Viel:2007mv}
therefore imply lower mass bounds in the WIMPless scenario, which
depend on the kinetic decoupling temperature $T_{\text{kd}}$ for
different $\xiRH$.

In addition, another kind of dark matter lower mass limit follows from
the observation that the phase-space density of the dark matter
particles in galaxy halos cannot exceed the maximum value of the
phase-space density when the dark matter particles were in kinetic
equilibrium~\cite{Tremaine:1979we}. With knowledge of the WIMPless
dark matter's chemical potential, one can then apply these generalized
Tremaine-Gunn bounds from Ref.~\cite{Madsen:1991mz} to our scenario.
Detailed considerations of these effects are beyond the scope of this
study, but they may modify the lower bound on $m_X$ or, alternatively,
provide interesting astrophysical signals for the hidden dark matter
scenarios discussed here.

\section*{Acknowledgments}

We thank Manoj Kaplinghat and Jason Kumar for many helpful
discussions. JLF and HT are supported in part by NSF grants
PHY--0239817, PHY--0314712, and PHY--0653656, NASA grant NNG05GG44G,
and the Alfred P.~Sloan Foundation. HY is supported by NSF grant
PHY--0709742.




\begin{thebibliography}{99}

\bibitem{Dimopoulos:1989hk}
  S.~Dimopoulos, D.~Eichler, R.~Esmailzadeh and G.~D.~Starkman,
  Phys.\ Rev.\  D {\bf 41}, 2388 (1990).

\bibitem{Starkman:1990nj}
  G.~D.~Starkman, A.~Gould, R.~Esmailzadeh and S.~Dimopoulos,
  Phys.\ Rev.\  D {\bf 41}, 3594 (1990).

\bibitem{Lee:1956qn}
  T.~D.~Lee and C.~N.~Yang,
  Phys.\ Rev.\  {\bf 104}, 254 (1956).

\bibitem{Blinnikov:1982eh}
  S.~I.~Blinnikov and M.~Y.~Khlopov,
  Sov.\ J.\ Nucl.\ Phys.\  {\bf 36}, 472 (1982)
  [Yad.\ Fiz.\  {\bf 36}, 809 (1982)];
  Sov.\ Astron.\  {\bf 27}, 371 (1983)
  [Astron.\ Zh.\  {\bf 60}, 632 (1983)].

\bibitem{Hodges:1993yb}
  H.~M.~Hodges,
  Phys.\ Rev.\  D {\bf 47}, 456 (1993).

\bibitem{Berezhiani:1995am}
  Z.~G.~Berezhiani, A.~D.~Dolgov and R.~N.~Mohapatra,
  Phys.\ Lett.\  B {\bf 375}, 26 (1996)
  [arXiv:hep-ph/9511221].

\bibitem{Mohapatra:2000qx}
  R.~N.~Mohapatra and V.~L.~Teplitz,
  Phys.\ Rev.\  D {\bf 62}, 063506 (2000)
  [arXiv:astro-ph/0001362].

\bibitem{Berezhiani:2000gw}
  Z.~Berezhiani, D.~Comelli and F.~L.~Villante,
  Phys.\ Lett.\  B {\bf 503}, 362 (2001)
  [arXiv:hep-ph/0008105].

\bibitem{Mohapatra:2001sx}
  R.~N.~Mohapatra, S.~Nussinov and V.~L.~Teplitz,
  Phys.\ Rev.\  D {\bf 66}, 063002 (2002)
  [arXiv:hep-ph/0111381].

\bibitem{Ignatiev:2003js}
  A.~Y.~Ignatiev and R.~R.~Volkas,
  Phys.\ Rev.\  D {\bf 68}, 023518 (2003)
  [arXiv:hep-ph/0304260].

\bibitem{Foot:2003iv}
  R.~Foot,
  Phys.\ Rev.\  D {\bf 69}, 036001 (2004)
  [arXiv:hep-ph/0308254].

\bibitem{Berezhiani:2003wj}
  Z.~Berezhiani, P.~Ciarcelluti, D.~Comelli and F.~L.~Villante,
  Int.\ J.\ Mod.\ Phys.\  D {\bf 14}, 107 (2005)
  [arXiv:astro-ph/0312605].

\bibitem{Foot:2004wz}
  R.~Foot and R.~R.~Volkas,
  Phys.\ Rev.\  D {\bf 70}, 123508 (2004)
  [arXiv:astro-ph/0407522].

\bibitem{Green:1984sg}
  M.~B.~Green and J.~H.~Schwarz,
  Phys.\ Lett.\  B {\bf 149}, 117 (1984).

\bibitem{Gross:1984dd}
  D.~J.~Gross, J.~A.~Harvey, E.~J.~Martinec and R.~Rohm,
  Phys.\ Rev.\ Lett.\  {\bf 54}, 502 (1985).

\bibitem{Blumenhagen:2005mu}
  R.~Blumenhagen, M.~Cvetic, P.~Langacker and G.~Shiu,
  Ann.\ Rev.\ Nucl.\ Part.\ Sci.\  {\bf 55}, 71 (2005)
  [arXiv:hep-th/0502005].

\bibitem{Schabinger:2005ei}
  R.~Schabinger and J.~D.~Wells,
  Phys.\ Rev.\  D {\bf 72}, 093007 (2005)
  [arXiv:hep-ph/0509209].

\bibitem{Patt:2006fw}
  B.~Patt and F.~Wilczek,
  arXiv:hep-ph/0605188.

\bibitem{Strassler:2006im}
  M.~J.~Strassler and K.~M.~Zurek,
  Phys.\ Lett.\  B {\bf 651}, 374 (2007)
  [arXiv:hep-ph/0604261].

\bibitem{Georgi:2007ek}
  H.~Georgi,
  Phys.\ Rev.\ Lett.\  {\bf 98}, 221601 (2007)
  [arXiv:hep-ph/0703260].

\bibitem{Kang:2008ea}
  J.~Kang and M.~A.~Luty,
  arXiv:0805.4642 [hep-ph].

\bibitem{Chen:2006ni}
  X.~Chen and S.~H.~Tye,
  JCAP {\bf 0606}, 011 (2006)
  [arXiv:hep-th/0602136].

\bibitem{Kikuchi:2007az}
  T.~Kikuchi and N.~Okada,
  Phys.\ Lett.\  B {\bf 665}, 186 (2008)
  [arXiv:0711.1506 [hep-ph]].

\bibitem{MarchRussell:2008yu}
  J.~March-Russell, S.~M.~West, D.~Cumberbatch and D.~Hooper,
  JHEP {\bf 0807}, 058 (2008)
  [arXiv:0801.3440 [hep-ph]].

\bibitem{Hooper:2008im}
  D.~Hooper and K.~M.~Zurek,
  Phys.\ Rev.\  D {\bf 77}, 087302 (2008)
  [arXiv:0801.3686 [hep-ph]].

\bibitem{McDonald:2008up}
  J.~McDonald and N.~Sahu,
  JCAP {\bf 0806}, 026 (2008)
  [arXiv:0802.3847 [hep-ph]].

\bibitem{Kim:2008pp}
  Y.~G.~Kim, K.~Y.~Lee and S.~Shin,
  JHEP {\bf 0805}, 100 (2008)
  [arXiv:0803.2932 [hep-ph]].

\bibitem{Krolikowski:2008qa}
  W.~Krolikowski,
  arXiv:0803.2977 [hep-ph].

\bibitem{Gong:2008uz}
  Y.~Gong and X.~Chen,
  arXiv:0803.3223 [astro-ph].

\bibitem{Feng:2008ya}
  J.~L.~Feng and J.~Kumar,
  arXiv:0803.4196 [hep-ph].

\bibitem{Foot:2008nw}
  R.~Foot,
  arXiv:0804.4518 [hep-ph].

\bibitem{Feng:2008dz}
  J.~L.~Feng, J.~Kumar and L.~E.~Strigari,
  arXiv:0806.3746 [hep-ph].

\bibitem{Kolb:1985bf}
See, \eg,
  E.~W.~Kolb, D.~Seckel and M.~S.~Turner,
  Nature {\bf 314}, 415 (1985).

\bibitem{Foot:1996hp}
  R.~Foot and R.~R.~Volkas,
  Astropart.\ Phys.\  {\bf 7}, 283 (1997)
  [arXiv:hep-ph/9612245].

\bibitem{Steigman:1977kc}
G.~Steigman, D.~N.~Schramm and J.~E.~Gunn,
Phys.\ Lett.\  B {\bf 66} (1977) 202.

\bibitem{Steigman:2007xt}
For a recent review, see
G.~Steigman,
Ann.\ Rev.\ Nucl.\ Part.\ Sci.\  {\bf 57} (2007) 463
[arXiv:0712.1100 [astro-ph]].

\bibitem{Cyburt:2004yc}
R.~H.~Cyburt, B.~D.~Fields, K.~A.~Olive and E.~Skillman,
Astropart.\ Phys.\  {\bf 23} (2005) 313
[arXiv:astro-ph/0408033].

\bibitem{Fields:2006ga}
B.~Fields and S.~Sarkar,
arXiv:astro-ph/0601514.

\bibitem{Komatsu:2008hk}
E.~Komatsu {\it et al.}  [WMAP Collaboration],
arXiv:0803.0547 [astro-ph].

\bibitem{:2006uk}
    [Planck Collaboration],
  arXiv:astro-ph/0604069.

\bibitem{Hannestad:2006as}
S.~Hannestad, H.~Tu and Y.~Y.~Y.~Wong,
JCAP {\bf 0606} (2006) 025
[arXiv:astro-ph/0603019].


\bibitem{Hamann:2007pi}
J.~Hamann, S.~Hannestad, G.~G.~Raffelt and Y.~Y.~Y.~Wong,
JCAP {\bf 0708} (2007) 021
[arXiv:0705.0440 [astro-ph]].


\bibitem{Ichikawa:2008pz}
K.~Ichikawa, T.~Sekiguchi and T.~Takahashi,
arXiv:0803.0889 [astro-ph].


\bibitem{Popa:2008tb}
L.~A.~Popa and A.~Vasile,
JCAP {\bf 0806} (2008) 028
[arXiv:0804.2971 [astro-ph]].



\bibitem{Simha:2008zj}
V.~Simha and G.~Steigman,
JCAP {\bf 0806} (2008) 016 [arXiv:0803.3465 [astro-ph]].

\bibitem{Mangano:2005cc}
G.~Mangano, G.~Miele, S.~Pastor, T.~Pinto, O.~Pisanti and
P.~D.~Serpico,
Nucl.\ Phys.\  B {\bf 729} (2005) 221 [arXiv:hep-ph/0506164].

\bibitem{Coleman:2003hs}
T.~S.~Coleman and M.~Roos,
Phys.\ Rev.\  D {\bf 68} (2003) 027702
[arXiv:astro-ph/0304281].

\bibitem{Pukhov:1999gg}
A.~Pukhov {\it et al.},
arXiv:hep-ph/9908288.

\bibitem{Pukhov:2004ca}
A.~Pukhov,
arXiv:hep-ph/0412191.

\bibitem{Gondolo:1990dk}
P.~Gondolo and G.~Gelmini,
Nucl.\ Phys.\  B {\bf 360} (1991) 145.

\bibitem{Gondolo:2004sc}
P.~Gondolo, J.~Edsjo, P.~Ullio, L.~Bergstrom, M.~Schelke and E.~A.~Baltz,
JCAP {\bf 0407} (2004) 008  [arXiv:astro-ph/0406204].

\bibitem{Bertschinger:2006nq}
E.~Bertschinger,
Phys.\ Rev.\  D {\bf 74} (2006) 063509 [arXiv:astro-ph/0607319].

\bibitem{Loeb:2005pm}
  A.~Loeb and M.~Zaldarriaga,
  Phys.\ Rev.\  D {\bf 71}, 103520 (2005)
  [arXiv:astro-ph/0504112].

\bibitem{Profumo:2006bv}
  S.~Profumo, K.~Sigurdson and M.~Kamionkowski,
  Phys.\ Rev.\ Lett.\  {\bf 97}, 031301 (2006)
  [arXiv:astro-ph/0603373].

\bibitem{Hooper:2007tu}
D.~Hooper, M.~Kaplinghat, L.~E.~Strigari and K.~M.~Zurek,
Phys.\ Rev.\  D {\bf 76} (2007) 103515 [arXiv:0704.2558 [astro-ph]].

\bibitem{Narayanan:2000tp}
V.~K.~Narayanan, D.~N.~Spergel, R.~Dave and C.~P.~Ma,
Astrophys.\ J.\ {\bf 543} (2000) L103
[arXiv:astro-ph/0005095].

\bibitem{Viel:2005qj}
M.~Viel, J.~Lesgourgues, M.~G.~Haehnelt, S.~Matarrese and A.~Riotto,
Phys.\ Rev.\  D {\bf 71} (2005) 063534
[arXiv:astro-ph/0501562].

\bibitem{Abazajian:2005xn}
K.~Abazajian,
Phys.\ Rev.\  D {\bf 73} (2006) 063513
[arXiv:astro-ph/0512631].

\bibitem{Viel:2007mv}
M.~Viel, G.~D.~Becker, J.~S.~Bolton, M.~G.~Haehnelt, M.~Rauch and
W.~L.~W.~Sargent,
Phys.\ Rev.\ Lett.\  {\bf 100} (2008) 041304
[arXiv:0709.0131 [astro-ph]].

\bibitem{Tremaine:1979we}
S.~Tremaine and J.~E.~Gunn,
Phys.\ Rev.\ Lett.\  {\bf 42} (1979) 407.

\bibitem{Madsen:1991mz}
J.~Madsen,
Phys.\ Rev.\  D {\bf 44} (1991) 999.

\end{thebibliography}
\end{document}